%% file: main.tex
\documentclass[aps,superscriptaddress,twocolumn,showpacs]{revtex4}

\usepackage{amsmath}
\usepackage{latexsym}
\usepackage{amssymb}
\usepackage{graphicx}
\usepackage[colorlinks=true, citecolor=blue, urlcolor=blue]{hyperref}
\usepackage{float}
\usepackage{wrapfig}
\usepackage{mathtools}
\usepackage[dvipsnames]{xcolor}
\usepackage{amsfonts}
\usepackage{textcomp}
\usepackage{mathpazo}
\usepackage{comment}
\usepackage{blindtext}
\usepackage{pgfplots}
\usepackage{soul}

\usepackage{bbm}

\usepackage{xcolor}
\definecolor{myurlcolor}{rgb}{0,0,0.4}
\definecolor{mycitecolor}{rgb}{0,0.5,0}
\definecolor{myrefcolor}{rgb}{0.5,0,0}
\usepackage{hyperref}
\hypersetup{colorlinks,
linkcolor=myrefcolor,
citecolor=mycitecolor,
urlcolor=myurlcolor}

\usepackage[draft]{fixme}
\usepackage{amsmath,bbm}%,mathrsfs}
\usepackage{scrextend}
\usepackage{graphicx}
\usepackage{amsfonts}
\usepackage{amssymb}
\usepackage{amsmath,amssymb,amsthm,verbatim,graphicx,bbm}
\usepackage{mathrsfs}
\usepackage{mathtools}
\usepackage{color,xcolor,longtable}
\usepackage{mdframed}

\usepackage{algorithm}
\floatname{algorithm}{Protocol}
\usepackage{algcompatible}

\def\be{\begin{equation}}
\def\ee{\end{equation}}
\def\ben{\begin{eqnarray}}
\def\een{\end{eqnarray}}
\def\eea{\end{array}}
\def\bea{\begin{array}}

\newcommand{\Tr}[1]{\mathrm{Tr}#1}
\newcommand{\bei}{\begin{itemize}}
\newcommand{\eei}{\end{itemize}}
\newcommand{\ket}[1]{|#1\rangle}
\newcommand{\bra}[1]{\langle#1|}

\newcommand{\proj}[1]{\ket{#1}\!\bra{#1}}

\newcommand{\I}{\mathbbm{1}}

\renewcommand{\emph}[1]{\textbf{#1}}

\makeatletter
\newtheorem*{rep@theorem}{\rep@title}
\newcommand{\newreptheorem}[2]{%
\newenvironment{rep#1}[1]{%
 \def\rep@title{#2 \ref{##1}}%
 \begin{rep@theorem}}%
 {\end{rep@theorem}}}
\makeatother

\theoremstyle{plain}
\newtheorem*{thm*}{Theorem}
\newreptheorem{thm}{Theorem}

\theoremstyle{definition}

\theoremstyle{remark}

\usepackage[english]{babel}
\usepackage[T1]{fontenc}
\usetikzlibrary{calc, fit}
\usepgfplotslibrary{fillbetween}
\usepackage{tcolorbox}

\pgfplotsset{compat=1.10}
\input{journals}
\begin{document}

%\begin{document}

\title{One-sided DI-QKD secure against coherent attacks over long distances}
\author{Michele Masini}
\address{Laboratoire d'Information Quantique, Universit\'e libre de Bruxelles (ULB), Belgium}
\author{Shubhayan Sarkar}
\email{theundiscoveredshubhayan@gmail.com}
\address{Laboratoire d'Information Quantique, Universit\'e libre de Bruxelles (ULB), Belgium}
\address{Institute of Informatics, Faculty of Mathematics, Physics and Informatics,
University of Gdansk, Wita Stwosza 57, 80-308 Gdansk, Poland}

\begin{abstract}
Quantum Key Distribution (QKD) is a technique enabling provable secure communication but faces challenges in device characterization, posing potential security risks. Device-independent (DI) QKD protocols overcome this issue by making minimal device assumptions but are limited in distance because they require high detection efficiencies, which refer to the ability of the experimental setup to detect quantum states. It is thus desirable to find quantum key distribution protocols that are based on realistic assumptions on the devices as well as implementable over long distances. In this work, we consider a one-sided DI QKD scheme with two measurements per party and show that it is secure against coherent attacks up to detection efficiencies greater than 50.1\% specifically on the untrusted side. This is almost the theoretical limit achievable for protocols with two untrusted measurements. Interestingly, we also show that, by placing the source of states close to the untrusted side, our protocol is secure over distances comparable to standard QKD protocols. 
\end{abstract}

\maketitle

\section{Introduction}
Quantum Key Distribution (QKD) is a technique that allows two remote users to establish a secure key over an untrusted quantum channel. This can be used to encrypt communication. QKD allows two parties to perform secure communication relying only on the laws of quantum mechanics and trusting that the devices they use follow a specific mathematical model \cite{reviewQKD2009,reviewQKD2020}. QKD is now commercially available and used for various applications. However, the challenge with QKD is that ensuring the quantum devices behave as expected is difficult, potentially opening security loopholes in the protocol \cite{lydersen2010hacking,gerhardt2011full,jain2011device,bugge2014laser}. In response to these vulnerabilities, the concept of Device-Independent (DI) quantum key distribution emerges as a class of QKD protocols whose security can be proven with minimal assumptions about the devices \cite{ref:ab2007,Primaatmaja2023securityofdevice}. However, they are currently strongly limited in distance. This follows from the fact that they are much more sensitive than device-dependent protocols to photon losses, which in optical fibers width exponentially with the distance.

To address these challenges, one possible approach is to incorporate well-chosen additional assumptions and ensure their validity (see e.g. \cite{lo2012measurement,braunstein2012side,pawlowski2011semi,woodhead2015secrecy,woodhead2016semi,yin2014mismatch}). In this work, we consider an entanglement-based one-sided device-independent protocols (1SDI) scenario \cite{mayers2004unconditional,tomamichel2011uncertainty,branciard1sided,lucamarini2012device,Tomamichel_2013,Vallone_2014,gehring2015implementation,Xin:20}. We assume that the source and one party's device (Bob's) are completely untrusted, while the second party's device (Alice's) is trusted. In the standard 1SDI scenario, the trusted measurements are taken to be of a particular form. Here, we slightly tweak the scenario and assume that the trusted side measures two anti-commuting observables with a discrete number of outcomes, which is almost the standard 1SDI scenario but with no restriction on the dimension of the local state on the trusted side. In practice, in the two-input-output case, both assumptions are equivalent; however, from a numerical standpoint, it is significantly more convenient to impose this constraint than to enforce a specific measurement structure. Moreover, we also impose a standard assumption on trusted detectors, that the probability for a detector to click is independent of the basis choice (a.k.a. the fair-sampling assumption in the context of Bell experiments). This implies that rounds where Alice does not detect a photon can safely be discarded. As we do not make any assumption on the Hilbert space dimension or on the specific form of the measurements, we will say from now on that Alice's device is semi-trusted. From a real-world perspective, this scenario resembles a situation where a user needs to establish secure communication with a server. In this context, the device belonging to the server can be considered semi-trusted, as it is affiliated with a company that possesses the necessary resources and expertise to regularly test and validate their devices. Quantum steering has also been previously utilised for certification of states, measurements and randomness \cite{Alex, Supic, sarkar6,  sarkar7, Bharti,  sarkar8, Sarkar_2024, sarkar2024deviceindependentcertificationgmestates}.

The robustness against photon losses of various forms of 1SDI QKD protocols with discrete variables have previously been explored in \cite{branciard1sided,lucamarini2012device,Vallone_2014,ioannou2022receiver,Ioannou_2022}. However, all these protocols are based on different forms of post-selection, where the secret key is derived only from specific measurement outcomes, discarding data from other rounds. The use of protocols where certain outcomes coming from untrusted devices are discarded does not allow to prove security against the most general class of attacks in a device-independent scenario using the recently introduced Entropy Accumulation Theorem (EAT) framework of \cite{arnon2018practical,dupuis2020entropy,metger2022generalised}. Moreover, it was shown in \cite{de2016randomness,sandfuchs2023coherent} that protocols relying on post-selection are indeed sensitive to coherent attacks. 

In this work, we explore a generalised entanglement-based version of the well-known BB84 protocol \cite{BENNETT20147} and we demonstrate that, in the simple case where Bob represents the undetected quantum states as separate outcomes, our 1SDI QKD scheme is secure as long as Bob's measurement has a detection efficiency greater than $50.1\%$ which is well within the current experimental limits. Interestingly, this threshold roughly corresponds to the one where any two-untrusted-measurements protocol becomes insecure \cite{acin2016necessary}. Furthermore, as we do not perform any type of post-selection on the untrusted end, one can analyze our protocols using the generalised entropy-accumulation-theorem, allowing one to prove that our scheme is secure against coherent attacks. Finally, by placing a source of entangled photons close to Bob's detector, we estimate that our 1SDI QKD protocol can be effectively implemented over distances of roughly $247$ km, assuming that Bob's detectors and optical components have a total detection efficiency of $90\%$, that the visibility of the state prepared is of $99\%$, and that the detectors have a dark count rate of $p_d=10^{-6}$. All of these values have been achieved in a recent experiment on DI QKD \cite{diqkd-experiment}. For our purpose, we employ the recently developed numerical framework for lower bounding conditional Von Neumann entropy \cite{brown2021device}. In short, our work can be considered as an application of recently developed ideas to an old protocol (without post-selection) to improve its security to the theoretical limit against general attacks.

\section{1SDI scenario} We will begin by outlining our setting, the additional assumptions we make beyond the standard assumptions made in DI QKD \cite{Primaatmaja2023securityofdevice}, and the protocol considered in this work. 

\begin{figure}[t]
    \centering
    \includegraphics[width=\linewidth]{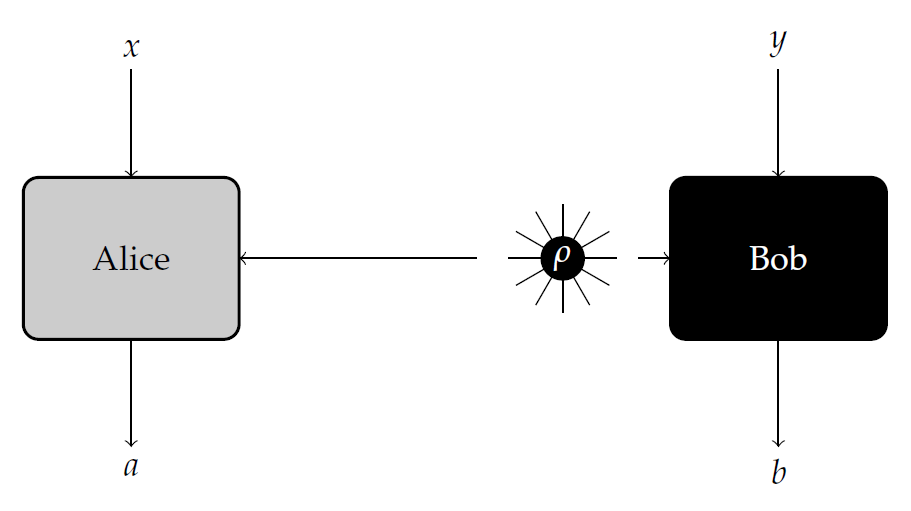}
    \caption{The scenario under consideration consists of an untrusted source preparing states $\rho$ and distributing them to both an untrusted measuring device (the black box), positioned near the source, and a semi-trusted measuring device (the gray box), situated far from the source.}
    \label{fig:steering}
\end{figure}

Within this work, we have two spatially separated observers, Alice and Bob, who share an entangled quantum state $\rho$ (Fig.~\ref{fig:steering}). The two parties choose the measurement settings $x=1,2$ and $y=1,2$ according to known distributions $p(x)$ and $p(y)$, and they perform measurements on their respective portions of the state. Alice obtains outcomes $a=1,2$, and $D=\checkmark,\perp$, while Bob obtains $b=1,2,\varnothing$. The measurement settings and $D$ are announced publicly. The rounds where $D=\perp$ are the ones where Alice did not detect a state and they are discarded (see Appendix A). Alice's and Bob's measurements are represented as $M_x=\{M_{a|x}\}$ and $N_y=\{N_{b|y}\}$, respectively where $M_{a|x}$, $N_{b|y}$ are POVMs. During the experiment, a fraction of the rounds are used to assess the security of the protocol by revealing publicly their outcomes to estimate the probability distribution $\vec{p}=\{p(a,b|x,y)\}$. We can write the observables of Alice and Bob as $A_{x}=M_{1|x}-M_{2|x}$ and $B_{y}=N_{1|y}-N_{2|y}$. The observables $A_1$, $A_2$ corresponding to Alice's measurements are trusted to obey the relation: 
\begin{equation}
    \{A_1,A_2\}=0. \label{eq:assumption}
\end{equation} 
This further implies that there exist some unitary $U_A$ acting on Alice's Hilbert space $\mathcal{H}_A$ which is isomorphic to $\mathbb{C}^2\otimes\mathcal{H}_{A''}$ ($\mathcal{H}_{A''}$ is some arbitrary Hilbert space) such that $U_AA_1U_{A}^{\dagger}=\sigma_z\otimes\I_{A''}, U_AA_2U_{A}^{\dagger}=\sigma_x\otimes\I_{A''}$ where $\sigma_z,\sigma_x$ are the Pauli $z,x$ observables. In contrast, Bob's measurement device is untrusted. 

An example of an honest implementation of the protocol is the case where Alice and Bob share a partially entangled state and perform the measurements of the observables $A_1=B_1=Z$ and $A_2=B_2=X$ and extract the secret key from the outcomes of the measurement of $Z$. The particular case where the state prepared is a maximally entangled state corresponds to an entanglement-based version of the BB84 protocol.

In a typical DI QKD scenario, the source of quantum states is placed in the middle between Alice and Bob and distributes two subsystems of an entangled state to the two parties. 
In this work, we consider a scenario (Fig.~\ref{fig:steering}) where the untrusted source is positioned near Bob's laboratory, a configuration previously explored in various contexts, e.g., \cite{branciard1sided,vallone2013einstein,gisin2010proposal,wollmann2016observation,zapatero2019long}. This arrangement minimizes losses during the transmission of the state to Bob's device, which is the one particularly sensitive to losses. On Alice's side, losses are less relevant as rounds with non-detection are simply discarded.  

\section{Key rate lower bounds}
Let us explicitly state the 1SDI protocol here;

\begin{enumerate}

\item \textbf{State distribution:}  
In each round, an untrusted source distributes a bipartite quantum state shared between Alice and Bob.

\item \textbf{Measurement choices:}  
Alice and Bob choose inputs $x,y \in \{1,2\}$ according to fixed distributions.

\item \textbf{Measurements:}  
Alice performs a trusted measurement $A_x$ and obtains $a \in \{1,2,\perp\}$.  
Bob performs an untrusted measurement and obtains $b \in \{1,2,\emptyset\}$.

\item \textbf{Sifting:}  
Rounds where Alice obtains no detection ($a=\perp$) are discarded.  
No post-selection is performed on Bob’s outcomes.

\item \textbf{Parameter estimation:}  
A subset of rounds is publicly revealed to estimate the statistics $p(a,b|x,y)$.

\item \textbf{Key generation:}  
The raw key is extracted from the outcomes corresponding to measurements $(x,y)=(1,1)$.

\end{enumerate}

\textbf{Key rate evaluation:}  
When the secret key is extracted from Alice's outcomes of the observables $A_1$ and Bob uses the outcomes of $B_1$ to reconstruct her string, the asymptotic key rate is given by 
\begin{equation}\label{eq:ent}
r_\infty \geq H(A_1|E) - H(A_1|B_1)
\end{equation}
where $H(A_1|E)$ is bounded using analytical and numerical techniques, and $H(A_1|B_1)$ is computed from the observed statistics. In line with standard QKD analyses, classical post-processing steps such as error correction and privacy amplification are implicitly accounted for in the Devetak–Winter rate \cite{devetak2005distillation,rennerrate}. Here, we decided to extract the key by using one-way public communication from Alice to Bob as we noticed that, with this choice, we can achieve better key rates.

The protocol described above is similar to the entanglement-based version of BB84 protocol with the exception that one of the measurement parties is untrusted. Moreover, we only discard the no-detection events on Alice's trusted side, which is a type of post-selection on trusted measurement devices. This is in contrast to Refs. \cite{branciard1sided,lucamarini2012device,Vallone_2014,ioannou2022receiver,Ioannou_2022} where post-selection is done on the untrusted end. It was shown in \cite{metger2022generalised} that post-selection on trusted devices satisfy the non-signalling constraints of generalised Entropy Accumulation Theorem (GEAT) and thus for such protocols, for example, in BB84 coherent attacks do not give any advantage to an attacker over collective attacks. Consequently, the 1SDI QKD protocol described above satisfies the non-signaling condition of the generalised Entropy Accumulation Theorem \cite{metger2022generalised} which allows us to assess their security in the asymptotic case by verifying the positivity of the Devetak-Winter rate \cite{devetak2005distillation,rennerrate}. 

It was already shown in \cite{dupuis2020entropy} that the standard entanglement-based BB84 protocol with one of the measurement devices being untrusted (our protocol), satisfies the conditions of EAT and then subsequently conditions of GEAT in \cite{metger2022generalised}. Here, we state the conditions with the full mathematical deatils adhered to \cite{dupuis2020entropy,metger2022generalised}. To apply the generalized EAT, the protocol must satisfy the following key requirements:

(i) Sequential structure (quantum Markov process):
The protocol must admit a description as a sequence of rounds, where in each round a quantum system is processed and produces classical outputs, and where the global evolution can be modeled as a concatenation of completely positive trace-preserving (CPTP) maps.
In our protocol, each round consists of state distribution followed by local measurements by Alice and Bob, producing outcomes $(a_r,b_r)$. Here, $r$ denotes the $r$-th round of the protocol. The entire protocol can thus be written as a sequential composition of maps acting on the joint system and the adversary’s side information, satisfying the structural requirement of EAT.

(ii) Well-defined classical outputs and conditioning registers:
Each round must produce classical data (e.g., inputs and outputs) that are used to define the entropy rate and the conditioning system.
In our setting, the classical registers consist of $(X_r,Y_r,A_r,B_r)$, where $X_r,Y_r$ denote the measurement choices and $A_r,B_r$ the outcomes. The raw key is extracted from a subset of these outputs (e.g., $(A_1,B_1)$), and the entropy accumulation is performed conditioned on the adversary’s quantum system.

(iii) No-signalling (generalized Markov condition):
This condition ensures that the evolution of the adversary’s side information does not depend on the internal memory of the devices, that is, future side information cannot reveal additional information about past outputs. In our setting, this condition is satisfied by construction of the protocol model. In particular, Alice’s device is trusted and memoryless, and her measurement operations act locally on the received quantum system without introducing hidden internal degrees of freedom that persist across rounds. Bob’s device and the source are treated as part of the adversarial system and are therefore fully included in Eve’s side information.
As a result, all inter-round correlations and memory effects are absorbed into the adversary’s system, and the honest part of the protocol (Alice’s operations) does not introduce additional hidden memory  influencing future rounds. 

(iv) No post-selection on the untrusted party (data retention):
A crucial requirement for EAT is that the sequence of rounds used for entropy accumulation is not adversarially filtered based on hidden information. In particular, post-selection on the untrusted party can invalidate the sequential structure needed for entropy accumulation.
In our protocol, while Alice discards rounds corresponding to no-detection events on her trusted side, no post-selection is performed on Bob’s outcomes, which originate from an untrusted device. All of Bob’s outputs, including no-click events, are retained in the data. This guarantees that the sequence of rounds entering the entropy accumulation is not biased by the adversary, and hence satisfies the data-processing requirements of generalized EAT.

 As one can observe from \eqref{eq:ent}, to find the asymptotic key rate, one needs to obtain the values of the conditional Von Neumann entropies $H(A_1|E)$ and $H(A_1|B_1)$.
As far as $H(A_1|B_1)$ is concerned, it can be estimated as
\begin{equation}
    H(A_1|B_1) = \sum_{a,b} p(a,b|1,1)\log_2\frac{p(a,b|1,1)}{p(b|1)}.
\end{equation} 
The above quantity can be directly inferred from the input-output statistics obtained through measurements of $A_1$ and $B_1$. In the above formula, the probabilities $p(a,b|1,1)$ are computed by post-selection on the trusted side; thus, only the detected rounds with Alice are utilised for obtaining the statistics and the key.

Let us now focus on $H(A_1|E)$. Here, we describe how to obtain a lower bound of $H(A_1|E)$ using two different techniques. The first one is analytical and based on the works of \cite{Woodhead2021deviceindependent,Masini2022simplepractical}, while the second one is numerical and based on \cite{brown2021device}. Here, we provide a sketch of both techniques while more detailed descriptions are provided in Appendix B and C.

Let us start with the analytical case. As shown in \cite{Masini2022simplepractical}, for measurements of Alice and Bob acting on two-dimensional spaces, $H(A_1|E)$ can be lower bounded as
\begin{equation}
H(A_1|E)\geq \phi \left(\langle A_1\rangle \right)
- \phi \left(\sqrt{\langle A_1\rangle^2+\langle \bar{A}_1\otimes B\rangle^2 } \right).\label{eq:bias_bound}
\end{equation} Here, $\phi(x)=h\left(\frac{1}{2}(1+x)\right)$, where $h(x)$ represents the binary entropy function. In this context, $\bar{A}_1$ is any observable that anti-commutes with the one employed for key extraction on Alice's side, while $B$ is a unitary operator, that is, $B^\dagger B=BB^{\dagger}=\mathbbm{1}$.
As described earlier, Alice's observables are trusted to be anti-commuting. Consequently, we can substitute $\bar{A}_1=A_2$ which allows us to easily bound the quantity $\langle \bar{A}_1\otimes B\rangle$ by estimating the correlator $\langle A_2\otimes B_2\rangle$. At this point, as we are in a two-input-two-output scenario, we can use Jordan's lemma \cite{jordanLemma} to make our bound valid for measurements of any dimension. This lemma asserts that all statistics in the two-input-two-output scenario can be derived by employing convex combinations of strategies involving Alice and Bob using qubits. In essence, if a lower bound is determined by assuming a qubit strategy, it must be convexified to account for the possibility of obtaining a lower bound through the mixture of strategies.  This ensures that one cannot obtain further lower bounds by convex mixing of two different strategies.
We show in Appendix C that the function appearing on the right-hand side of eq.~\eqref{eq:bias_bound} is convex in $\langle A_1\rangle$ and $\langle A_2\otimes B_2\rangle$. Hence, we can say that our bound holds for states and measurements of any dimension. 

The application of Jordan's lemma necessitates the restriction that the statistics employed to bound the adversary's information are derived only from two inputs per party with only two possible outcomes. Therefore, for the specific rounds where we estimate $\langle A_2\otimes B_2\rangle$, we need to modify the protocol by deterministically mapping Bob's non-detection outcomes ($\varnothing$) to one of his other possible outcomes. 

Alternatively, to compute $H(A_1|E)$, we can adopt a second approach focused on numerical techniques based on the NPA (Navascués-Pironio-Ac\'in) hierarchy of semi-definite programs (SDP) \cite{Navascues_2008,Pironio_npa}. We will denote it as BFF (Brown-Fawzi-Fawzi) technique. As shown in \cite{brown2021device}, the quantity $H(A_1|E)$ can be lower bounded by the solution to a noncommutative polynomial optimization problem. In our case, we include into the optimization problem a further polynomial constraint (given by eq.~\eqref{eq:assumption}) that will force Alice's measurements to anti-commute (Appendix B). 

The latter analysis can be used to analyze more general scenarios which include both the case where Bob maps his non-detection outcomes into one of the other outcomes and the case where he keeps it as a separate outcome. Moreover, it is possible to constrain the information of Eve not only by taking into account the correlators $\langle A_2\otimes B_2\rangle$ and $\langle A_1\rangle$, but using the full probability table $p(a,b|x,y)$. 

\textit{Reference experiment.---} To simulate the performance of an experiment, we will assume an honest implementation of the protocol where we consider a reference state prepared by the source and reference measurements performed by both parties that generate the probability distribution $\vec{p}$. We assume that the source prepares the two-qubit entangled states of the form
\begin{equation}\label{eq:state}
\ket{\psi(\theta)} = \cos\theta\ket{00}+\sin\theta\ket{11}.
\end{equation}
Alice measures the Pauli operators $A_1=Z$ and $A_2=X$. Since she is semi-trusted, she can discard the non-detection events due to which her measurements will not be affected by detection losses. For Bob's measurements, we include losses by evolving his measurements as
\begin{align}
    N_{1|y}(\eta) &= \eta N_{1|y}+(1-\eta)\mathbbm{1}, \label{eq:loss1} \\
    N_{2|y}(\eta) &= \eta N_{2|y}, \label{eq:loss21}
\end{align} where $N_{b|y}$ are the projective measurements that Bob performs in the ideal case, and $\eta$ represents the transmittance or detection efficiency. To obtain the optimal key rates we will maximize the lower bound to the key rate over the parameter $\theta$ of the state $\ket{\psi(\theta)}$  \eqref{eq:state} for every value of $\eta$ taken into account. For the ideal measurements, we choose $B_1=Z$ as this will make Bob's measurements used for the secret key correlated to Alice's ones, and $B_2=X$ as this will maximize the correlator $\langle A_2\otimes B_2\rangle$ and thus also the value of the conditional entropy in eq.~\eqref{eq:bias_bound} (we expect the case where we use BFF to be analogous). 

\begin{figure}[t]
\centering
\includegraphics[width=\linewidth]{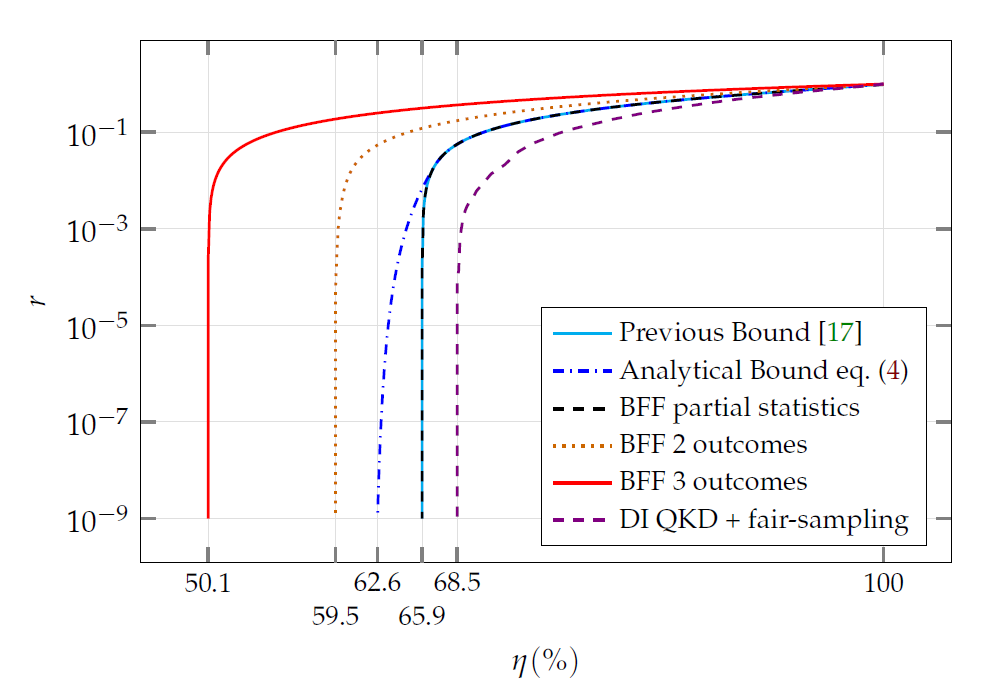}
\caption{Plot of the key rate as a function of Bob's detection efficiency computed using different techniques as explained in the main text. As Alice is trusted, only the detected rounds with Alice are utilised for obtaining the statistics and the key.}\label{fig:rates}
\end{figure}

We now plot the lower bound to the key rates in Fig.~\ref{fig:rates} obtained using the above-mentioned techniques.
Firstly, we display the bound of \cite{branciard1sided}. This bound was obtained for a protocol that includes post-selection on the untrusted party and the key rate was estimated using a different technique which allowed to prove security only for the case of memoryless devices. Notice that, as we show in Appendix C, the bound of \cite{branciard1sided} can be obtained also under the same assumptions considered in this work without performing post-selection on Bob's side and, thus, allowing us to prove security without assuming memoryless devices. In particular, the conditional entropy of Alice can be bounded by
\begin{equation}
    H(A_1|E)\geq 1-\phi(\langle A_2\otimes B_2\rangle).
\end{equation} Hence, using the measurements described above and partially entangled states (eq.~\eqref{eq:state}), the key rate becomes
\begin{equation}
    r_\infty\geq 1-h\left(\frac{1}{2}(1+\eta\sin(2\theta))\right)-(1-\eta)h(\cos^2(\theta)),
\end{equation} and its maximum is at $\theta=\pi/4$ for all $\eta$, meaning that the optimal state in this case is always a maximally entangled state. 

Secondly, we plot the bound obtained using eq.~\eqref{eq:bias_bound}. In this case, we obtained an improvement that comes from the use of partially entangled states together with the inclusion of the information on the bias $\langle A_1\rangle$ in the bound. 

Additionally, we include in Fig.~\ref{fig:rates} the key rates for three different scenarios computed using the BFF technique. Firstly, we have the case where Bob maps his non-detection events into one of his other outcomes, and we only constrained the mean value $\langle A_2\otimes B_2\rangle$. In this case, displayed with a black dashed line in Fig.~\ref{fig:rates}, we recovered a lower bound on the key rate which is almost identical to the one identified in \cite{branciard1sided}. We maximized the key rate over the angle of the partially entangled state for all values of $\eta$, but, as in the analytical case, the optimal angle was always $\theta\simeq\pi/4$, i.e., a maximally entangled state. Using the same technique, we could recover also the bound of eq.~\eqref{eq:bias_bound} by constraining the mean values $\langle A_2\otimes B_2\rangle$ and $\langle A_1\rangle$ and maximizing over the partially entangled states. Anyway, for this case, the closer we got to the threshold the more we had to increase the level of the NPA hierarchy and the number of Gauss-Radau coefficients (see Appendix B for the definition), and this was possible up to slightly less than 64\%, where the key rate was of the order of $10^{-5}$ and the precision of the SDP solver used was not enough to provide reliable results. We did not encounter this problem in the first case as the key rate drops more quickly as a function of $\eta$.

Secondly, we analyzed the same protocol using the full probability distribution $\vec{p}_{\mathrm{ref}}$ as a constraint on our problem. This procedure led to an improvement of the key rate which was positive up to a detection efficiency of $59.5\%$ as we can see from the orange dotted line. In this case, the use of optimal angles $\theta$ led to a small improvement. 

Lastly, we computed the key rate using full statistics in the scenario where Bob keeps undetected photons as a separate outcome (red line). For this last case, the optimal $\theta$ was again the one of a maximally entangled state for all $\eta$. Here, we could show the positivity of the key rate up to $50.1\%$. Notice that at $\eta=50\%$ Bob's measurements become jointly measurable, and thus no secret key can be extracted under our assumptions \cite{acin2016necessary,masinijm}. 

\begin{figure}[t]
    \centering
    \includegraphics[width=\linewidth]{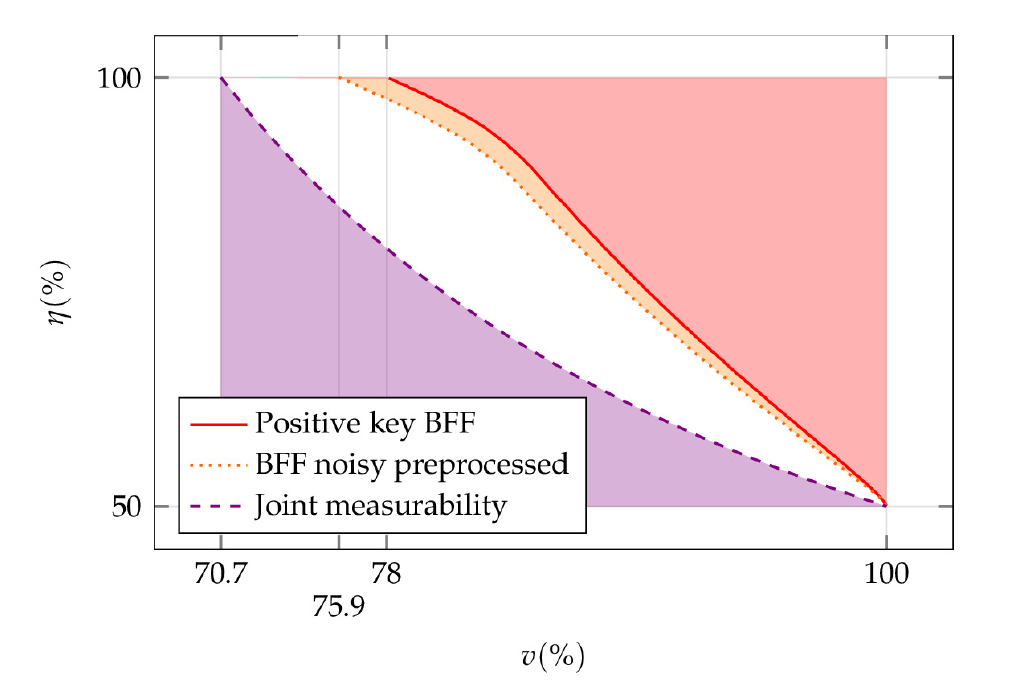}
    \caption{Regions where the key rate becomes zero as a function of visibility and detection efficiency. The region where the BFF 3 outcome protocol with partially entangled states is secure is above the red line, the case where we include also noisy preprocessing is above the orange dotted line, while the purple line denotes the boundary below which Bob's observables are jointly measurable, rendering any protocol insecure in that region.}
    \label{fig:etav}
\end{figure}

For the sake of comparison, we include in Fig.~\ref{fig:rates} the plot of an analogous setting for the case of a DI scenario where we make a fair sampling assumption on Alice's side. This case is explained in Appendix F. 

We will now study the robustness of our technique in the presence of depolarizing noise acting on the initial state together with photon losses. The reference statistics are obtained by adding depolarizing noise to the state generated by the source as
\begin{equation}
    \rho = v\proj{\psi(\theta)} +\frac{1-v}{4}\mathbbm{1},
\end{equation} where $v$ is denoted as visibility. We show in Fig.~\ref{fig:etav} the region in the space of parameters $\eta$ and $v$ where the key rate is positive. This analysis is conducted in two scenarios. One where we optimize the key rate only over the angle of the partially entangled state and another where we additionally perform optimal noisy pre-processing \cite{noisyp}, a technique where Alice randomly flips her outcomes during key generation rounds with a probability $q$. The key rate was computed using BFF with full statistics and keeping Bob's undetected photons as a separate outcome. We noticed that for visibilities smaller than one, the use of partially entangled states (and of noisy pre-processing) leads to an improvement also for the BFF case where the three outcomes of Bob are kept separate. Moreover, we include in the plot a curve from \cite{masinijm} which displays the zone where the evolution of $X$ and $Z$ measurements of Bob under depolarizing noise and losses are jointly measurable. When Bob's measurements are jointly measurable, the correlations between Alice and Bob are unsteerable. As a result, no secure key can be extracted in this region regardless of the initial state prepared, and regardless of the measurements performed by the semi-trusted party.
One can see in Fig.~\ref{fig:etav} that, for non-unit values of the visibility, a region exists wherein the key rate is non-positive, and the noisy versions of $X$ and $Z$ observables are not jointly measurable. This suggests that refining our protocol or improving its analysis might enhance its resilience against such noise. Specifically, exploring variations of our protocol involving secret key extraction from both Alice's measurements and using the BFF technique with higher levels of the NPA hierarchy may prove insightful. However, recall that the joint measurability of Bob's measurements is only a necessary condition for the protocol to be insecure. For non-unit values of visibility, stricter conditions may be necessary.

\section{Theoretical distance estimate of implementability} Finally, in order to estimate the distance at which our protocols can be implemented, we will deploy a model where Alice's and Bob's devices observe dark counts. We describe our model in Appendix E. It is crucial to note that the noise model we are about to use is regarded as part of the untrusted quantum channel that evolves the untrusted state $\rho$. Therefore, as it is typically assumed in standard QKD protocols, dark counts are not included in the model of Alice's semi-trusted device, ensuring that her measurements remain anti-commuting. For convenience, we will evolve Alice's ideal POVMs to compute the probability distribution resulting from our model as this is equivalent to evolving the state $\rho$.

Dark counts are essentially random clicks generated by detectors resulting from thermal fluctuations inside the detectors. These are of critical importance particularly when operating at low detection efficiencies, as their frequency becomes noteworthy over the fraction of rounds that have not been discarded under the fair sampling assumption.

Let us provide Bob's measurement operators. We choose to categorize all the events where Bob experiences zero or double clicks as a separate outcome $\varnothing$. As explained in Appendix E, his POVMs evolve as
\begin{align}
    N_{1,2|y}(\eta_B,p_d) &= \eta_B(1-p_d) N_{1,2|y}+(1-\eta_B)p_d(1-p_d)\mathbbm{1}, \\ 
    N_{\varnothing|y}(\eta_B,p_d) &= \big(p_d\eta_B +(1-\eta_B)(p_d^2+(1-p_d)^2)\big)\mathbbm{1},
\end{align} 
where $N_{b|y}$ are the projective measurements that Bob performs in the ideal case. 

\begin{figure}[t]
    \centering
    \includegraphics[width=\linewidth]{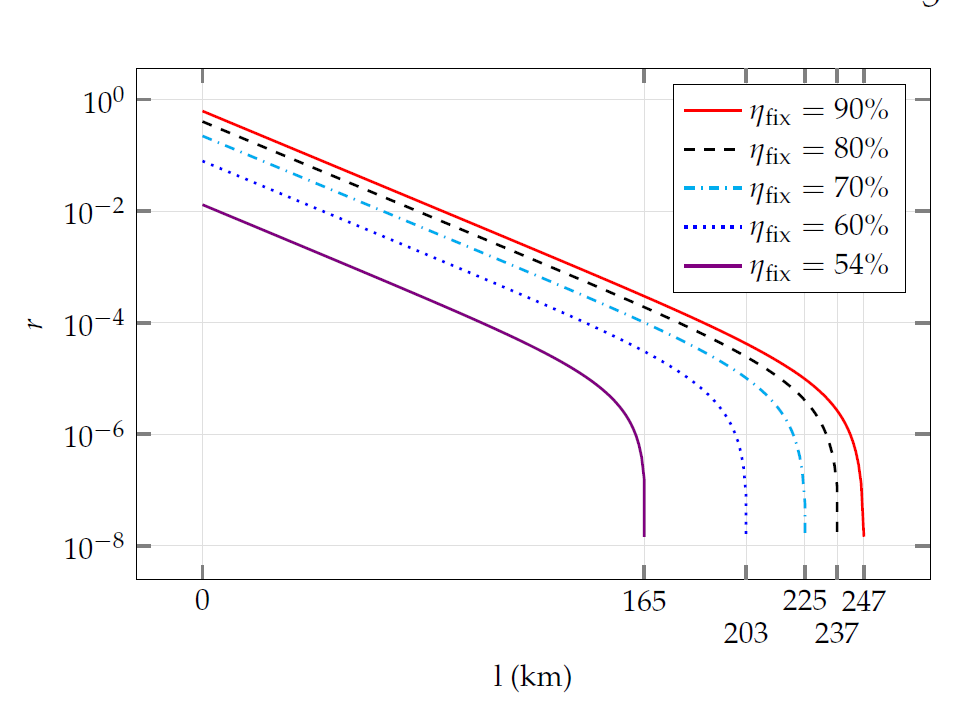}
    \caption{Plot of the key rate as a function of the distance in a model that includes photon losses, depolarizing noise, and dark counts.}
    \label{fig:distance}
\end{figure}

On Alice's side, we choose to discard all events where her detectors do not click or where they both click. As we discard these events, we will need to renormalize her POVMs dividing them by the probability of having only one click. We show in Appendix E that the measurement operators of Alice $M_{1|x}(\eta_A,p)$ can be expressed in terms of the ideal ones $M_{1|x}$ as
\begin{equation}
    M_{1|x}(\eta_A,p_d) = \frac{ \eta_A(1-p_d) M_{1|x} +(1-\eta_A)p_d(1-p_d)\mathbbm{1}}{1-p_d\eta_A -(1-\eta_A)(p_d^2+(1-p_d)^2)},
\end{equation} and $M_{2|x}(\eta_A,p_d)=\mathbbm{1}-M_{1|x}(\eta_A,p_d)$. 

For the reference statistics, we will consider the values of visibility and dark counts of the most advanced experiment we are aware of, which employs an entanglement-based scheme \cite{diqkd-experiment}. Specifically, we take $v=0.99$ and $p_d=10^{-6}$. As far as the detection efficiency is concerned, we will take into account a fixed component $\eta_\text{fix}$ due to the efficiency of the detectors and of the other optical components, and a variable component that accounts for optical fiber losses based on its length $l$, given by $\eta_\text{fiber}=10^{-\alpha\cdot l/10}$. Here, we will use the typical value for telecom wavelength of $\alpha=0.2\, dB/km$ (see e.g. \cite{zapatero2019long,acin2016necessary}). Since the source is placed very close to Bob's laboratory, we can ignore the losses arising due to the fibre. Consequently, his efficiency will be $\eta_B=\eta_\text{fix}$, while Alice's one is $\eta_A(l)=\eta_\text{fix}\cdot 10^{-\alpha \cdot l/10}$. Given these parameters of the setup, in Fig.~\ref{fig:distance} we plot the lower bound to the asymptotic key rate as a function of the distance. To show the number of key bits generated in each actual round of the protocol (and not only in the ones that have been retained), the key rates in Fig.~\ref{fig:distance} have been multiplied by the probability of retaining a round which is $1-p_d\eta_A -(1-\eta_A)(p_d^2+(1-p_d)^2)$. To compute the key rates, we used the BFF method where Bob's third outcomes are kept separate. We provide the plot for different values of $\eta_\text{fix}$. For each specific $\eta_\text{fix}$ and $l$ value, we optimized the key rate by adjusting the partially entangled state and the noisy pre-processing. 

\section{Discussions}
We applied to a 1SDI scenario the most recent methods to derive lower bounds on the conditional Von Neumann entropy. We provided lower bounds with both analytical and numerical methods and showed that our analytical results can be recovered with the numerical method of \cite{brown2021device}. We showed that the first QKD protocol ever introduced, the BB84 protocol, can be proven secure in an entanglement-based 1SDI scenario with up to 50.1\% detection efficiency on the untrusted side. This result is almost optimal as no protocol employing two untrusted measurements can achieve security below a 50\% detection efficiency threshold. As per GEAT \cite{metger2022generalised}, in protocols with finite $n$ rounds, the key rate per round is lower bounded by the asymptotic rate minus a correction term, which is of the order of $1/\sqrt{n}$. Since the BFF technique respects this bound \cite{brown2021device}, our result can straightaway be adapted to the finite-round regime, thus making it more apt for practical scenarios. Furthermore, we observed that, by increasing the number of Gauss-Radau coefficients, we can get closer to this critical threshold. It remains an open question whether our result is optimal also in terms of depolarizing noise. It will also be interesting to obtain key rates using the above approach when one can trust Alice only up to some error \cite{Sarkar_2023}. Another approach to tackle this problem would be to utilise the semi-device-independent protocol suggested in \cite{Woodhead_2016}, where the only assumption of the trusted side is the dimension of the local state. However, one still has to verify whether the protocol can be made secure against general attacks.

%It is important to point out here that we model the 1SDI-QKD setup using a source generating single-photon pairs and thus photon losses can be modelled in a particular way as in Eq. \eqref{eq:loss21}. This is one of the most natural ways to model photon losses when the sources are generating single-photons \cite{PNS1}. This is also imposed when we are extracting the key using the above scheme. However, realistic setups might be utilising attenuated lasers that might open up the possibility of photon number splitting attacks \cite{PNS2, PNS3}. An interesting problem might be to analyse this scenario in the 1SDI-QKD framework and find ways to overcome this problem to make 1SDI-QKD more feasible from a practical perspective. One possible ways might be to utilise decoy states as done for BB84 protocol \cite{PNS4}.

Note added: These results are also later confirmed in independent works for obtaining key rates using local Bell tests \cite{routedBell1, Le_Roy_Deloison_2025}. These can be considered as a robust version of our protocol, that is, the measurements are not fully trusted, but up to some error. However, both these works need to assume that one of the components, a "switch" which decides which way the signals would travel, to be trusted.

\begin{acknowledgments}
    We extend our gratitude to Stefano Pironio for his extensive discussions and valuable insights regarding our work. His contributions have been fundamental in enhancing the quality and depth of our research. We thank Thomas Van Himbeeck for discussions on the possibility of extending the security of our work to the finite-size case. We thank Erik Woodhead for developing an open-source Julia library to compute NPA relaxations, and Abhishek Mishra for his work in extending it. We thank Marco Avesani for the discussions on estimating the distance of implementability of our work. We acknowledge funding from the VERIqTAS project within the QuantERA II Programme that has received funding from the European Union’s Horizon 2020 research and innovation programme under Grant Agreement No 101017733 and the F.R.S-FNRS Pint-Multi programme under Grant Agreement R.8014.21, from the European Union’s Horizon Europe research and innovation programme under the project "Quantum Security Networks Partnership" (QSNP, grant agreement No 101114043),  from the F.R.S-FNRS through the PDR T.0171.22, from the FWO and F.R.S.-FNRS under the Excellence of Science (EOS) programme project 40007526, from the FWO through the BeQuNet SBO project S008323N, from the Belgian Federal Science Policy through the contract RT/22/BE-QCI and the EU ``BE-QCI" programme. S.S. also acknowledges the National Science Centre, Poland, grant Opus 25, 2023/49/B/ST2/02468.
Funded by the European Union. Views and opinions expressed are however those of the authors only and do not necessarily reflect those of the European Union, which cannot be held responsible for them.
\end{acknowledgments}

\noindent
\textbf{Author contributions}  M.M. and S.S. contributed equally to the manuscript.
\\
\textbf{Data availability}  The code used to compute the key rates numerically can be found in the GitHub folder \cite{github}.
\\
\textbf{Declarations}
\\
\textbf{Ethics approval and consent to participate} The manuscript is not submitted to another journal. The submitted work is original and has not been published elsewhere.
\\
\textbf{Consent for publication} The authors are responsible for correcting the statements provided in the manuscript.
\\
\textbf{Competing interests} The authors declare no competing interests.

%\bibliography{refs}
\input{ref.bbl}

\newpage

\appendix

\onecolumngrid
\section{Alice's undetected photons}

\begin{figure}[H]
    \centering
    \includegraphics[width=.5\linewidth]{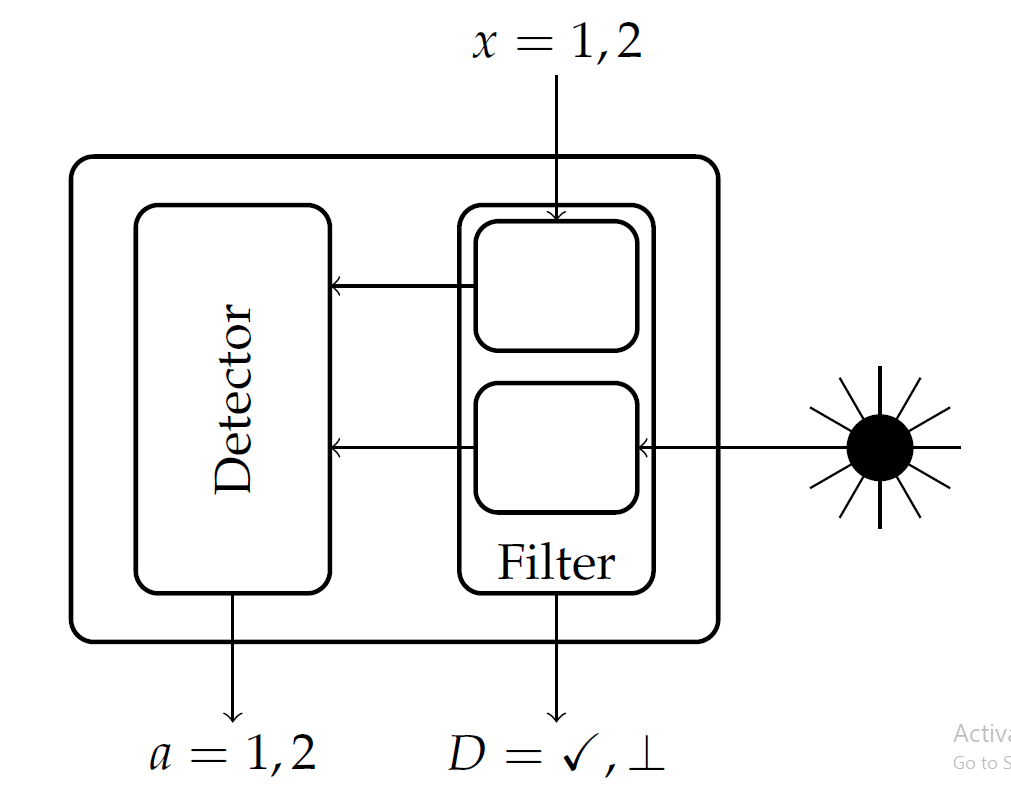}
    \caption{A model describing Alice's device under the fair sampling assumption. A filter returns an output $D=\checkmark,\perp$ which is determined independently by the input $x$ and the state $\rho$. The detector performs a lossless measurement only if the filter returned $D=\checkmark$.}
    \label{fig:filter}
\end{figure}

Let us describe in this section the model describing Alice's device. To account for no-detection events on Alice's side, we model her semi-trusted device inspired by \cite{Orsucci2020howpostselection}. Her device is composed of a filter and an ideal detector (Fig.~\ref{fig:filter}). The filter outputs a flag $D=\checkmark,\perp$ based on whether her device produced a click or not. The filter is composed of two parts acting independently on the classical and quantum input. When the filter receiving the quantum input accepts the round, the device of Alice produces a click. If her device does not click, the round of the protocol gets discarded. If the filter produces a click $D=\checkmark$, Alice's device performs one of the two ideal anti-commuting measurements according to the random input $x$ and obtains outcomes $a$. Making this assumption on Alice's device is equivalent to performing an ideal measurement with unit efficiency.  In contrast, if Bob fails to detect an incoming state, he stores a separate outcome as $b=\varnothing$.

\section{Brown-Fawzi-Fawzi bounds}\label{ap:bff}

In this section, we provide the non-commutative polynomial optimization problem that we used to lower bound the conditional Von Neumann entropy to verify the security of our protocols. Our lower bound is a simple extension of the work of \cite{brown2021device} to a steering scenario.

Let $t_1,\dots,t_m$ and $w_1,\dots,w_m$ be the nodes and weights of an $m$-point Gauss-Radau quadrature on $(0,1]$ where we fix $t_m=1$. These coefficients can be computed efficiently in terms of Legendre polynomials \cite{gauss_radau}. Moreover, let 
\begin{equation}
    \alpha_i=\frac{3}{2}\min \left\{ \frac{1}{t_i},\frac{1}{1-t_i}\right\}.
\end{equation}

We have that 
\begin{equation}
    H(A_1|E) \geq \sum_{i=1}^{m-1}\frac{w_i}{t_i \ln(2)}+ \sum_{i=1}^{m-1}\frac{w_i}{t_i \ln(2)}O_i,
\end{equation} where $O_i$ are the solutions to the following polynomial optimization problems
\begin{align}
    \text{inf}\quad &\sum_a \bra{\psi}M_{a|1}(Z_{a,i}+Z^*_{a,i}+(1-t_i)Z_{a,i}^*Z_{a,i})+t_iZ_{a,i}Z^*_{a,i}\ket{\psi} \\
    \text{s.t.}\quad & \bra{\psi}M_{a|x}N_{b|y}\ket{\psi}=p(a,b|x,y),\\
    & \sum_a M_{a|x}=\sum_b N_{b|y}=\mathbbm{1},  \\
    & M_{a|x}\geq 0,\quad N_{b|y}\geq 0,  \\
    & Z_{a,i}^*Z_{a,i}\leq\alpha_i,\quad Z_{a,i}Z^*_{a,i}\leq\alpha_i,  \\
    & [M_{a|x},N_{b|y}]=[M_{a|x},Z^{(*)}_{a,i}]=[Z_{a,i}^{(*)},N_{b|y}]=0,  \\
    & (M_{1|1}-M_{2|1})(M_{1|2}-M_{2|2})+ (M_{1|2}-M_{2|2})(M_{1|1}-M_{2|1})=0 \label{eq:constr-anti}.
\end{align} In the last constraint (eq.~\eqref{eq:constr-anti}) of the optimization problem, we introduced the anti-commutation constraint. Note that we expressed w.l.o.g. $A_x=M_{1|x}-M_{2|x}$. The probabilities $p(a,b|x,y)$ are the ones observed during the experiment or, in our case, computed from the simulations.

In the case where we perform noisy preprocessing, the measurement operators $M_{a|1}$ in the objective function need to be replaced with
\begin{align}
    M_{1|1} &\rightarrow (1-q)M_{1|1}+qM_{2|1}, \\
    M_{2|1} &\rightarrow (1-q)M_{2|1}+qM_{1|1}.
\end{align}

The latter problem is a non-commutative polynomial optimization problem and it can be relaxed to a hierarchy of semi-definite programs \cite{Navascues_2008}. The lower bound that we obtain from the solution of this problem is increasingly tighter as a function of the level of the hierarchy and as a function of the number of points of the Gauss-Radau quadrature. For the computations performed in this work in the case where Alice's observables are assumed to anti-commute, we kept the level of the localizing matrices of the NPA hierarchy fixed to $1$ and used $m=15$ in the Gauss Radau quadrature. The level of the principal moment matrix was slightly bigger than one in order to make sure that all the terms that appear in the localizing moment matrices were also present in the principal one. Additionally, our analysis revealed that for the problems examined in this article, incorporating localizing matrices representing the constraints $Z_{a,i}^* Z_{a,i}\leq\alpha_i$ and $Z_{a,i}Z^*_{a,i}\leq\alpha_i$ did not improve the lower bounds on the conditional Von Neumann entropy. Consequently, we opted for simpler constraints $\bra{\psi}Z_{a,i}^*Z_{a,i}\ket{\psi}\leq\alpha_i$ and $\bra{\psi}Z_{a,i}Z^*_{a,i}\ket{\psi}\leq\alpha_i$, which still establish a valid lower bound on the conditional Von Neumann entropy.

Finally, in the case where we analyzed a DI protocol with a fair sampling assumption on Alice's side, we did not include any localizing matrix. This comes from two reasons. First, we did not need to enforce anti-commutativity and, second, as before, we used the simpler constraints $\bra{\psi}Z_{a,i}^*Z_{a,i}\ket{\psi}\leq\alpha_i$ and $\bra{\psi}Z_{a,i}Z^*_{a,i}\ket{\psi}\leq\alpha_i$. However, in this scenario, achieving satisfactory results required the use of a principal moment matrix at level $2+ABZ$.

\section{Analytical bounds}\label{ap:analytical}

In this section, we will prove extensively the analytical lower bound on the key rate described in the main text.

We will bound the quantity $H(A_1|E)$ as a function of input-output statistics derived from measurements of $A_{1,2}$ and $B_{1,2}$ that only have two possible outcomes. In this way, we can use Jordan's lemma \cite{jordanLemma} and reduce our problem to deriving a convex lower bound on the conditional Von Neumann entropy between Alice and Eve in the case where Alice and Bob's systems are two-dimensional. 

If the secret key is extracted from the outcomes of the measurement of the two-outcomes observable $A_1$ on Alice's side, the conditional Von Neumann entropy of Alice conditioned on Eve can be bounded by \cite{Woodhead2021deviceindependent}
\begin{equation}
    H(A_1|E)\geq 1- \phi \left(|\langle \bar{A}_1\otimes B\rangle |\right), \label{eq:simpleH}
\end{equation} where $B$ is any unitary observable on Bob's subspace, $\bar{A}_1$ is an observable on Alice's subspace that anticommutes with $A_1$, and $\phi(x)=h(1/2(1+x))$ where $h(x)$ is the binary entropy function. Under the same assumptions, a second bound that we can use \cite{Masini2022simplepractical} is
\begin{equation}
    H(A_1|E)\geq \phi \left(\langle A_1\rangle \right) \\
    - \phi \left(\sqrt{\langle A_1\rangle^2+\langle \bar{A}_1\otimes B\rangle^2} \right),\label{eq:bias_bound_ap}
\end{equation}

At this point, we need to find a lower bound on $\langle \bar{A}_1\otimes B \rangle$ as a function of the input-output statistics of Alice and Bob's devices such that it is independent of what is the form of the measurements performed by Bob. 

For the steering scenario, we can choose to bound the correlator directly. By assuming that $\bar{A}_1=A_2$, and choosing $B=B_2$, we can evaluate $\langle A_2\otimes B_2\rangle$ from the input-output statistics. In this case, using eq.~\eqref{eq:bias_bound_ap}, our bound becomes of the type
\begin{equation}
    H(A_1|E)\geq \phi \left(\langle A_1\rangle \right) 
    - \phi \left(\sqrt{\langle A_1\rangle^2+\langle A_2\otimes B_2\rangle^2} \right).\label{eq:bias_bound2}
\end{equation} Notice that for the case of eq.~\eqref{eq:simpleH} applied to an honest implementation where Alice and Bob measure Pauli operators $Z,X$ on a maximally entangled state undergoing losses and depolarizing noise, we obtain the same lower bound of \cite{branciard1sided} just by assuming the anti-commutativity of Alice's measurements and making a fair sampling assumption on her side.

Finally, in order to use Jordan's lemma, we will need to prove that the bounds obtained are convex in the variables $x=\langle A_2\otimes B_2\rangle$ and $z=\langle A_1\rangle$.

The case of eq.~(\ref{eq:simpleH}) is straightforward since the binary entropy function is a concave function. Let us now prove the convexity for the case of eq.~(\ref{eq:bias_bound_ap}). 

We start with the direct case. We need to show that, for the function
\begin{equation}
    f(z,x) = \phi(z)-\phi\left(\sqrt{z^2+x^2}\right),
\end{equation}
the hessian matrix is such that
\begin{equation}
    \mathbf{H}(f(z,x)) = \begin{pmatrix}
         & \frac{d^2}{dz^2}f(z,x) & \frac{d^2}{dz\,dx}f(z,x) \\ 
         \\
         & \frac{d^2}{dz\,dx}f(z,x) & \frac{d^2}{dx^2}f(z,x)
    \end{pmatrix} \geq 0.
\end{equation} Being a $2\times 2$ matrix, the positivity is ensured by the positivity of the diagonal elements and of the determinant. We have
\begin{equation}
    \frac{d^2}{dz^2}f(z,x) = \frac{x^2}{\ln (2)}  \left( \frac{x^2-1+2z^2}{(z^2+x^2)(1-x^2-z^2)(1-z^2)} +\frac{\text{arctanh}\! \left(\sqrt{z^2+x^2}\right)}{\left(x^2+z^2\right)^{3/2}}\right). \label{eq:dzq}
\end{equation} we can easily notice that, since $z^2\leq 1$, the positivity of eq.~(\ref{eq:dzq}) is ensured by the condition $z^2+x^2\leq 1$ which is proven in appendix~\ref{sec:condition}.

The second diagonal element is
\begin{equation}
    \frac{d^2}{dx^2}f(z,x) = \frac{1}{\ln (2)} \left( \frac{x^2}{(z^2+x^2)(1-x^2-z^2)} +z^2\frac{\text{arctanh}\! \left(\sqrt{z^2+x^2}\right)}{\left(x^2+z^2\right)^{3/2}}\right), \label{eq:dxq}
\end{equation} whose positivity is ensured by the same conditions as before. Finally, the determinant is
\begin{multline}
    \frac{d^2}{dz^2}f(z,x)\frac{d^2}{dx^2}f(z,x)-\left(\frac{d^2}{dx\,dz}f(z,x)\right)^2 =  \\
    \frac{x^2}{(\ln(2))^2(z^2+x^2)^2(1-x^2-z^2)(1-z^2)} \bigg(\sqrt{z^2+x^2} \, \text{arctanh}\,\sqrt{z^2+x^2} 
     -(z^2+x^2)  \bigg), \label{eq:detH}
\end{multline} whose positivity is ensured by $z^2\leq 1$, by $z^2+x^2\leq 1$, and by $\text{arctanh} (x)\geq x$.

We conclude that eq.~\eqref{eq:bias_bound2} is a valid lower bound under the assumptions of anti-commutativity between $A_1$ and $A_2$ and under a fair sampling assumption on Alice's side. Notice that Alice's Hilbert space dimension does not need to be bounded.

\section{Domain of the correlations} \label{sec:condition}
In this section, we prove the condition 
\begin{equation}\label{eq:border}
    \langle A\rangle^2+ \langle\bar{A}\otimes B\rangle^2\leq 1,
\end{equation} which is required in Section~\ref{ap:analytical}.

    Let us begin by considering a state $\rho_{AB}$ and then purifying it to $\ket{\psi_{ABE}}$  such that $\rho_{AB}=\Tr_E\ \psi_{ABE}$ where $E$ denotes the ancillary system. Now, any general state $\ket{\psi_{ABE}}\in\mathbbm{C}^2\otimes\mathcal{H}_{BE}$ can be written as
    \begin{equation}\label{genstate1}
    \ket{\psi_{ABE}}=\sum_{i=0,1}\lambda_i\ket{i}_{A}\ket{e_i}_{BE}
    \end{equation}
    where $\lambda_i\geq0$, $\sum_{i=0,1}\lambda_i^2=1$ and $\ket{e_i}_{BE}$ are normalised but in general not orthogonal. Now, evaluating the left-hand side of eq.~\eqref{eq:border} by plugging in the state \eqref{genstate1}, we obtain
    \begin{equation}
        \langle Z\rangle^2+ \langle X\otimes B\rangle^2=\left(\lambda_0^2-\lambda_1^2\right)^2+4\lambda_0^2\lambda_1^2\ \left(\mathrm{Re}\langle e_0|B|e_1\rangle\right)^2
    \end{equation}
    where we used the fact that $B$ is Hermitian. As $B$ is unitary, we have that $-1\leq\mathrm{Re}\langle e_0|B|e_1\rangle\leq1$ using which we arrive at
    \begin{equation}
         \langle Z\rangle^2+ \langle X\otimes B\rangle^2\leq\left(\lambda_0^2-\lambda_1^2\right)^2+4\lambda_0^2\lambda_1^2=\left(\lambda_0^2+\lambda_1^2\right)^2=1.
    \end{equation}
    This completes the proof.

\section{A model for dark counts and losses}\label{ap:dark}
In this section, we present a model for computing the measurement operators of Alice and Bob, taking into account photon loss and dark counts. 
\begin{figure}[H]
    \centering
  \includegraphics[width=.5\linewidth]{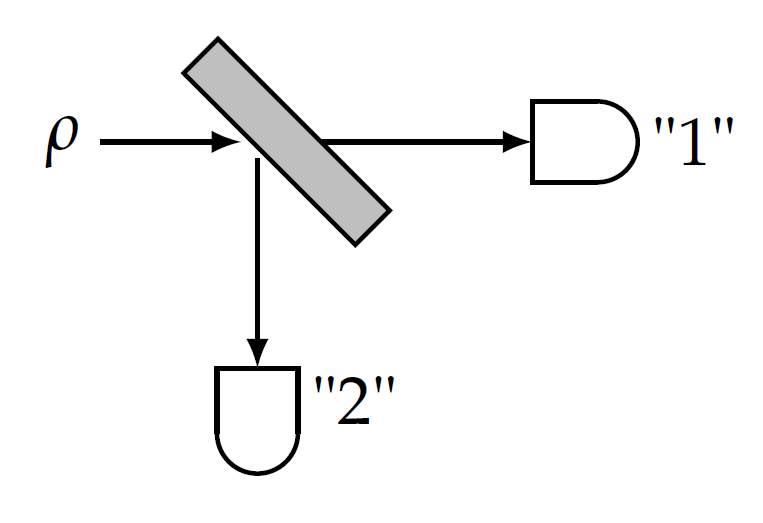}
    \caption{Schematic representation of an ideal detector in the absence of losses and dark counts. A polarizing beam splitter divides the incoming photonic state into two paths. At the two output ports of the beam splitter, we place two detectors and each of them corresponds to one of the two possible outcomes. Outcomes "1" and "2" correspond to two orthogonal polarizations of the photon.}
    \label{fig:detector}
\end{figure}

In Fig.~\ref{fig:detector}, we represent an ideal detector. Let us initially address photon loss. When a photon arrives at the measurement device (with probability $\eta$), it encounters a polarizing beam splitter, which then directs it to one of two detectors based on its polarization state. If the photon is lost (with probability $1-\eta$), then neither detector clicks.

Let us now assume that the detectors of Alice and Bob exhibit also a dark count behavior, with each of the two possible outcomes having a probability of $p_d$ of registering a spurious click in every round of the protocol. 
Our setup does not allow us to differentiate whether the detection was due to a genuine photon detection, or a dark count.

Let us delineate the four possible detection outcomes.
\begin{enumerate}
    \item \textbf{Single click in the first detector:} This outcome indicates that only the detector "1" clicked. It could arise from the detection of a photon with a specific polarization or a dark count in the first detector.
    \item \textbf{Single click in the second detector:} This outcome indicates that only the detector "2" clicked. %It might be due to the detection of a photon with the orthogonal polarization or a dark count in the second detector.
    \item \textbf{Double click:} It occurs when both detectors register a click simultaneously. This can arise from both detectors experiencing dark counts, or one detecting a photon while the other registers a dark count.
    \item \textbf{No click:} When neither detector registers a click, suggesting the photon was lost and no dark counts occurred.
\end{enumerate}

Let us now evaluate the POVMs related to each of these four events. We will take into account the ideal POVMs $M_{1}$ and $M_{2}$ of the photon being detected by the first and second detectors, respectively, the efficiency $\eta$ of the photon's arrival at the detectors, and the probability $p_d$ of a dark count occurring in each detector.

Assuming the photon arrives at the measurement device, four possible events can occur. We list them below. %in table~\ref{tab:photon_arrival_outcomes}.
\begin{table}[H]
\centering
\begin{tabular}{|l|l|l|}
\hline
\textbf{Event} & \textbf{POVM} & \textbf{Outcome Type} \\
\hline
Photon at first detector (no dark count at second) & $\eta \times (1 - p_d) \times M_1 $ & Single click in the first
\\ \hline
Photon at second detector (no dark count at first) & $ \eta \times (1 - p_d) \times M_2 $ & Single click in the second \\
\hline
Photon at first detector
(dark count at second) & $\eta \times p_d \times M_1 $ & Double click \\ 
\hline
Photon at second detector
(dark count at first) & $\eta \times p_d \times M_2 $ & Double click \\
\hline
\end{tabular}
%\caption{POVMs of detection events with photon arrival, considering dark counts and categorized by outcome type.}
\label{tab:photon_arrival_outcomes}
\end{table}

For the scenario where the photon does not arrive, we have four cases which we list below. %in table~\ref{tab:no_photon_arrival_outcomes}.

\begin{table}[H]
\centering
\begin{tabular}{|l|l|l|}
\hline
\textbf{Event} & \textbf{POVM} & \textbf{Outcome Type} \\
\hline
No dark count in either detector & $(1-\eta)\times (1 - p_d)^2 \times \mathbbm{1} $ & No click \\ \hline
Dark count in the first detector only & $(1-\eta)\times p_d \times (1 - p_d) \times \mathbbm{1} $ & Single click in the first \\
\hline Dark count in the second detector only & $(1-\eta)\times (1 - p_d) \times p_d \times \mathbbm{1} $ & Single click in the second \\ \hline
Dark counts in both detectors & $(1-\eta)\times p_d^2 \times \mathbbm{1} $ & Double click \\
\hline
\end{tabular}
%\caption{POVMs of detection events without photon arrival, categorized by outcome type.}
\label{tab:no_photon_arrival_outcomes}
\end{table}

To compute the measurement operators of Bob, given the ideal measurement operators $N_{1|x}$ for outcome 1 and $N_{2|x}$ for outcome 2, we need to consider the probabilities of each event as previously discussed. We obtain in this way four measurement operators $N_{b|y}(\eta_B,p_d)$ corresponding to the four outcome types, where $\eta_B$ is the detection efficiency at Bob's side.

\begin{align}
    N_{1|y}(\eta_B,p_d) &= \eta_B(1-p_d) N_{1|y}+(1-\eta_B)p_d(1-p_d)\mathbbm{1}, \\ 
    N_{2|y}(\eta_B,p_d) &= \eta_B(1-p_d) N_{2|y}+(1-\eta_B)p_d(1-p_d)\mathbbm{1}, \\ 
    N_{\text{Dc}|y}(\eta_B,p_d) &= \big(p_d\eta_B +(1-\eta_B)p_d^2\big)\mathbbm{1}, \\
    N_{\text{Nc}|y}(\eta_B,p_d) &= (1-\eta_B)(1-p_d)^2\mathbbm{1}.
\end{align} 

For simplicity, we will group Bob's no clicks and double clicks into a single outcome, obtaining in this way
\begin{equation}
        N_{\varnothing|y}(\eta_B,p_d) = \big(p_d\eta_B +(1-\eta_B)(p_d^2+(1-p_d)^2)\big)\mathbbm{1}.
\end{equation}

Let us finally focus on Alice's measurements. Her ideal POVMs are denoted as $M_{a|x}$. Given the assumption that the probability of a click in Alice's detector is independent of the basis choice, we are allowed to discard events where both her detectors do not click or where they both click. As we discard these events, we will need to renormalize Alice's POVMs dividing them by the probability of having only one click. Consequently, the measurement operators of Alice $M_{a|x}(\eta_A,p)$ can be expressed in terms of the ideal ones $M_{1|x}$ as
\begin{equation}
    M_{1|x}(\eta_A,p_d) = \frac{ \eta_A(1-p_d) M_{1|x} +(1-\eta_A)p_d(1-p_d)\mathbbm{1}}{1-p_d\eta_A -(1-\eta_A)(p_d^2+(1-p_d)^2)},
\end{equation} and $M_{2|x}(\eta_A,p_d)=\mathbbm{1}-M_{1|x}(\eta_A,p_d)$. Here, $\eta_A$ is the detection efficiency on Alice's side.

\section{1SDI QKD without anti-commutation assumption}
In this section, we will describe how we computed the key rate for the case where we do not assume that Alice's observables anticommute.
In this case, Alice performs two measurements and she is modeled as described in Section~I, while Bob measures three different observables each with three possible outcomes. The ideal observables of Alice and Bob are
\begin{align}
    A_x &= \cos(\alpha_x)Z+\sin(\alpha_x)X, \\
    B_y &= \cos(\beta_y)Z+\sin(\beta_y)X.
\end{align} 

\begin{figure}[t]
    \centering
    \includegraphics[width=.5\linewidth]{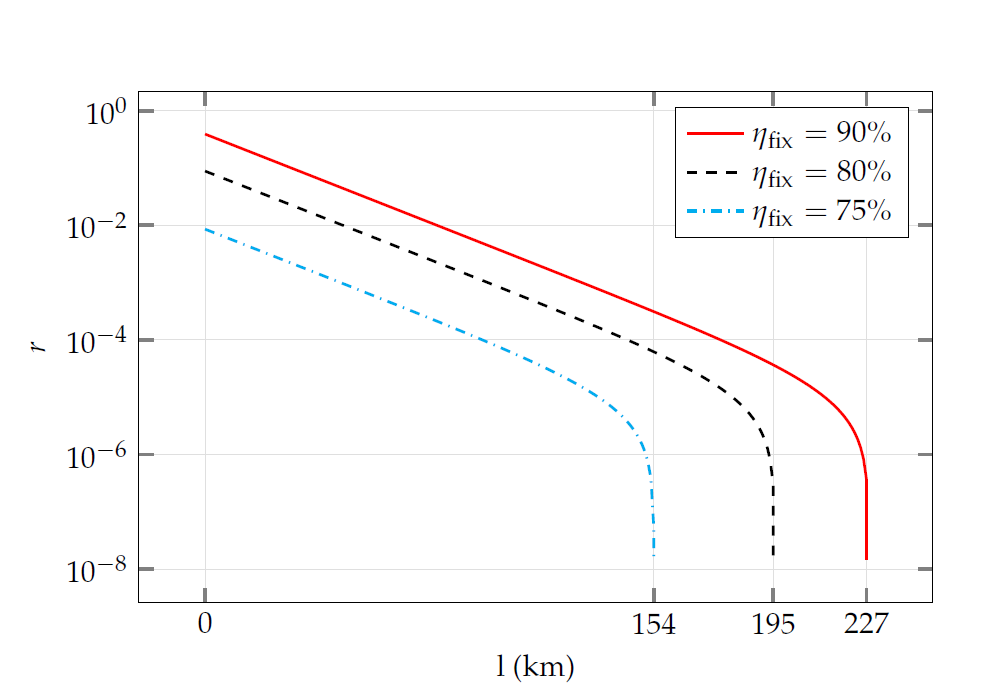}
    \caption{Plot of the key rate as a function of the distance in a model that includes photon losses, depolarizing noise, and dark counts in the case where we do not assume that Alice's measurements anti-commute.}
    \label{fig:DIdistance}
\end{figure}

We apply the same noise model described in the main text, hence, Bob's measurement operators are evolved according to \begin{align}
    N_{1|y}(\eta) &= \eta N_{1|y}+(1-\eta)\mathbbm{1}, \\
    N_{2|y}(\eta) &= \eta N_{2|y}, \label{eq:loss2}
\end{align} where $N_{b|y}$ are the projective measurements that Bob performs in the ideal case, and $\eta$ represents the transmittance or detection efficiency. Alice's measurement operators remain the ideal ones due to the fair-sampling assumption. Here, Alice has only two possible outcomes as she discards the undetected photons, while Bob keeps his three outcomes separate. The secret key is extracted from the outcomes of $A_1$, while Bob uses the outcomes of $B_3$ to guess Alice's ones. Moreover, we use a technique denoted as \textit{noisy pre-processing} \cite{noisyp}, where Alice randomly flips her outcomes during key generation rounds with a probability $q$. We did not use this technique for the other cases from Fig.~2 of the main text as our best result allowed us to obtain a 50.1\% threshold which is already very close to optimal. We finally compute the key rates with BFF using the full probability distribution as a constraint. We optimize the key rate heuristically at each value of $\eta$ over the parameters $\theta$, $\alpha_1$, $\alpha_2$, $\beta_1$, $\beta_2$, $\beta_3$, $q$ and obtain a threshold of $68.5\%$.

Finally, we include in Fig.~\ref{fig:DIdistance} a plot where we estimate the achievable distance in the case of a protocol where we do not assume that Alice's observables anti-commute and Bob makes three measurements as explained before. Similarly to the case reported in the main text, we computed the key rates using the BFF method where Bob's third outcomes are kept separate, and we fixed $v=0.99$ and $p_d=10^{-6}$. In this case, for each specific $\eta_\text{fix}$ value, we optimized the key rate by adjusting the parameters $\theta$, $\alpha_1$, $\alpha_2$, $\beta_1$, $\beta_2$, $\beta_3$, $q$ for a distance of $l=0$, and then applied the same parameters to compute the key rate across all other distances. This was due to the fact that computing a key rate in this setting requires a higher level of the NPA hierarchy (compared to the case where we assume anti-commutativity of Alice's measurements) and optimizing over the parameters in question for each value of $l$ would require a too large computational time. Let us remark however that the optimization for each value of $l$ performed for the plot in the main text resulted only in a few kilometers of improvement compared to the case where we optimize only for $l=0$ and apply the same parameters for each $l$. This is due to the fact that the key rate drops quickly to zero when $\eta_A$ and $p_d$ reach a similar order of magnitude.
We can see that we can obtain similar distances as in the case of the main text, but Bob's device needs greater detection efficiencies. In particular, for $\eta_\text{fix}\lesssim 70\%$ this type of protocol is not secure, while the one where we assume that Alice's observables anticommute can still be proven secure. This requirement can be inconvenient in a client-server scenario where Bob is a user with limited resources.

%\section{References}

\end{document}

%% file: journals.tex
\def\bsmf{Bull.\ Soc.\ Math.\ Fr.}%
\def\natcomm{Nat.\ Commun.}%
\def\pra{Phys.\ Rev.\ A}%
\def\prl{Phys.\ Rev.\ Lett.}%
\def\prsa{Proc.\ R.\ Soc.\ A}%

%% file: ref.bbl
\providecommand{\noopsort}[1]{}\providecommand{\singleletter}[1]{#1}%